 \documentclass[final,3p,times]{elsarticle}

\usepackage[dvips]{color}                 
\usepackage{graphicx}              
\usepackage{multirow}                     
\usepackage{amssymb}
\usepackage{amsmath}
\usepackage{xspace}                       
\ifx\feynVersion\AnswerYes
   \usepackage{feynmp}                    
\fi

\newcommand{\3}{\ss}

\newcommand{\non}{\nonumber}

\newcommand{\half}{\frac{1}{2}}

\newcommand{\ii}{\mathrm{i}}
\newcommand{\dd}{\mathrm{d}}

\newcommand{\deint}[2]{\dd^{#1}\! #2\;}

\newcommand{\kv}{\vec{k}}
\newcommand{\pv}{\vec{p}}
\newcommand{\qv}{\vec{q}}

\newcommand{\LambdaNoPion}{\ensuremath{\Lambda_{\pi\hskip-0.4em /}}}

\newcommand{\calH}{\mathcal{H}} \newcommand{\calK}{\mathcal{K}}

\journal{Few-Body Systems}

\begin{document}

\begin{frontmatter}


 \title{Triton Photodisintegration with Effective Field Theory}
\author[1]{H. Sadeghi\corref{cor1} \fnref{label2}}
\ead{H-Sadeghi@araku.ac.ir}
\author[2]{S. Bayegan}
\address[1]{Department of Physics, University of Arak, P.O.Box 38156-879, Arak, Iran.}
\address[2]{Department of Physics, University of Tehran, P.O.Box 14395-547, Tehran, Iran.}
\fntext[label2]{Tel:+988614173400, Fax:+988614173406}
\cortext[cor1]{Corresponding author}

\begin{abstract}
Effective field theory (EFT) has been recently used for the
calculation of neutron-deuteron radiative capture at very low
energies. We present here the use of EFT to calculate the two-body
photodisintegration of the triton, considering the three-body force.
The calculated cross section shows sharply rising from threshold to
maximum about 0.88 mb at $\sim$ 13 MeV and decreasing slightly to
about  0.81 mb at $\sim$ 19 MeV, in corresponding to the
experimental results. Our results are in good agreement with the
experimental data and the other modern realistic two- and
three-nucleon forces AV18/UrbanaIX potential calculation.
\end{abstract}


\end{frontmatter}


\section{Introduction} \label{introduction}
The total photodisintegration cross sections of the three-body
nuclei are important quantities for understanding electromagnetic
processes in nuclei. The three-nucleon systems involving
nucleon-deuteron radiative capture and photodisintegration of $^3$He
and $^3$H have been investigated in experimental and theoretical
works over the past decades.

In the experimental investigations, a few measurements of triton
photodisintegration have been done. More recently, including of the
two- and three-body reactions, photodisintegration cross sections
are measured at low energies~\cite{Faul}. Mono energetic photons are
used and neutrons are detected, in this experiment. The obtained
$^3H(\gamma, n){^2}H$ cross section shows sharply rising from
threshold to the maximum of  0.9 mb at $\sim$ 12 MeV and decreasing
only slightly to  0.8 mb at $\sim$ 19 MeV. The results are the
two-body breakup cross sections for $^3$H~\cite{Faul}. The same
shape for the $^3$He cross section has been reported but with the
lower magnitude.

During the past several years, different theoretical approaches have
been used in order to investigate the cross sections of the
nucleon-deuteron radiative capture and photodisintegration of $^3$He
and $^3$H from the very low energy up to 100 MeV.  The  main aim of
these theoretical studies has been the calculations of the bound
state and scattering wave functions in order to be used in matrix
elements of the electromagnetic current operators and ultimately
testing the underlying two, three body interactions as well as
currents with certain degree of complexity for the three-nucleon
systems. In many theoretical treatments, by solving the
three-nucleon Faddeev equations based on simple finite rank forces,
it is found that the interacting three-nucleon continuum is crucial
for understanding the process in three-body
reactions~\cite{ref.Barbour&Phillips}. Sandhas and his collaborators
investigated many theoretical treatments of photo-effect in $^3$H
and $^3$He three-nucleon systems by employing the realistic
nucleon-nucleon forces exclusive process at ~0-40 MeV
energies\cite{Sandhas1,Sandhas2}. For covering a larger energy
range, the inclusive two- and three-body breakup cross sections have
been studied by Efros et al.~\cite{Efros}. Recently, the Bonn group
used the modern nucleon-nucleon forces. In these works, the current
was restricted to the dominant $E_1$ multipole~\cite{ref.Bonn} or to
the $E_1$ and $E_2$ multipoles~\cite{ref.Bonn.2}. They compared the
Faddeev approach and a hyperspherical harmonic expansion together
with the Lorentz integral transform method using AV18 together with
Urbana IX.

For the $p$-$d$ and $n$-$d$ radiative capture, a magnetic dipole M1
transition is a dominant contribution at the very low energy, in the
plane wave(Born)approximation~\cite{friar}. In such investigations,
the authors employed Faddeev calculations, with the inclusion of
three-body forces and meson exchange current(MEC).  More recently, a
very good agreement reached with the calculation of the total
three-nucleon photodisintegration cross section with two quite
different approaches~\cite{Trento}. Rather detailed investigation of
such processes has been performed by Viviani et~al.~\cite{Viviani}.
Recent calculation using manifestly gauge invariant currents reduced
the spread~\cite{Viviani2}, but the result including three-body
currents, $0.558$ mb, still over-predicts the cross-section by 10\%.
The $V_{\rm UCOM}$ potential leads to a very similar description of
the cross section as the AV18 interaction, considering the Urbana IX
three-body force for photon energies $45\le \omega \le 120$ MeV,
while larger differences are found close to threshold~\cite{Bacca}.

During the last few years, nuclear pionless EFT has been applied to
the two-, three-nucleon systems
~\cite{kaplan,Beane,3stooges_boson,3stooges_doublet,
doubletNLO,4stooges,chickenpaper,20} and recently developed pionless
EFT is particularly suited to the high order precision calculation.
A recent model independent neutron-proton radiative capture
calculation has been done, using EFT to predict a theoretical
uncertainty 1\% for center of mass energies $E\leq 1 MeV$
~\cite{Rupak}. At these energies, which are relevant for big-bang
nucleosynthesis, isovector magnetic transitions $M_1$ and isovector
electric transitions $E_1$ give the dominant contributions. In past
years, we have calculated the cross section of $nd\rightarrow
{^3H}\gamma$ process~\cite{Sadeghi1,Sadeghi2,Sadeghi3}. No new
three-nucleon forces are needed in order to achieve cutoff
independent results of neutron-deuteron radiative capture process up
to next-to-next-to leading order(N$^2$LO). The three body parameters
will be fixed by the triton binding energy and Nd scattering length
in the triton channel.

In this paper, we study the two-body photodisintegration of the
triton $\gamma{^3H}\rightarrow nd$ using pionless EFT and insertion
of the three-body force, up to N$^2$LO. The evaluated cross section
has been compared with experimental data and the total cross section
of the three-nucleon photodisintegration calculation of the modern
realistic two- and three-nucleon forces AV18/UrbanaIX potential
models. Close agreement between the available experimental data and
the calculated cross section is reached. We demonstrate convergence
and cutoff independence order by order in the low energy expansion.

The paper is organized as follows: in Section~\ref{section:Theorey},
we briefly review theoretical framework for calculating the Faddeev
integral equation, Lagrangians for the reaction involved in our
consideration, the relevant diagrams, three-body forces and cutoffs
dependence. Comparisons of our results with the corresponding
experimental and theoretical data are given in
Section~\ref{section:comparison}. Summary and conclusions follow in
Section~\ref{section:conclusion}.

\section{ Theoretical framework}
\label{section:Theorey}

In the present work we confine ourselves with the electric and
magnetic dipole approximation for the transition operators. These
transitions are known to make most of the total cross section at
least for the low energy region where the pionless EFT theory is
most reliable. The photodisintegration reaction, $\gamma
{^3H}\rightarrow nd$, in these energies occur predominantly via
electric and magnetic, specially electric-dipole transition. The
triton photodisintegration cross section is then given
by~\cite{Trento}:
\begin{equation}
\sigma_{\gamma {^3H} \rightarrow nd}(E_\gamma)  =  (2\pi)^2
(\frac{e^2}{\hbar c})M(E_\gamma)E_\gamma , \label{eq:photo}
\end{equation}
where $E_\gamma$ is the incident photon energy and $M(E_\gamma)$
will be calculated from electric- and magnetic-dipole transition
amplitudes. These transition amplitudes have been prepared by
considering of the Faddeev equation for final nd-scattering and for
the initial triton bound state system to some order (e.g. LO, NLO
and N$^2$LO). We then take these Faddeev amplitudes and folded with
the photon-interaction with nucleons when the photon kernel is
expanded to the same order.

\begin{figure}[!htp]
\begin{center}
\includegraphics*[width=.5\textwidth]{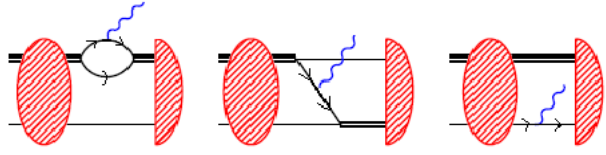}
\vspace*{-0pt} \caption{ The Faddeev equation for $Nd$-scattering
and diagrams for adding photon-interaction to the Faddeev equation
contribution to $E_1$ amplitude.Wavy line shows photon and thick
line is propagator of the two intermediate auxiliary fields $D_s$
and $D_t$; $\calK$: propagator of the exchanged nucleon, see
eq.(2.2).} \label{fig1}
\end{center}
\end{figure}

The Lagrangians relevant to photodisintegration of the triton can be
divided into:

 (a) Lagrangian that contributes to
nucleon-deuteron scattering. The derivation of the integral equation
describing neutron-deuteron systems has been discussed before, see
e.g.~\cite{3stooges_doublet,griesshammer,20,Sadeghi2}. Only the
results will be used here.  Two mixed configuration in the
three-nucleon system are exist. The spin-triplet auxiliary deuteron
field $D_t$ combines, depending on their spin directions, with the
nucleon $N$ to total quartet channel spin ($S=\frac{3}{2}$) or
doublet channel($S=\frac{1}{2}$). The Faddeev equation is introduced
as a coupled equation and is given by~\cite{20}:
\begin{eqnarray}
  \label{eq:doubletpw}
  \lefteqn{{t^{(l)}_{d,tt}\choose t^{(l)}_{d,ts}}(E;k,p)=2\pi\;
  \calK^{(l)}(E;k,p)\;{1\choose -3}}\\ \non
  &&-\;\frac{1}{\pi}\int\limits_0^\infty\deint{}{q} q^2\;
    \calK^{(l)}(E;q,p)\;\begin{pmatrix}1&-3\\-3&1\end{pmatrix}
    \begin{pmatrix}D_t(E-\frac{q^2}{2M},p)&0\\0&D_s(E-\frac{q^2}{2M},p)
    \end{pmatrix}
    {t^{(l)}_{d,tt}\choose t^{(l)}_{d,ts}}(E;k,q)\;,
  \non
\end{eqnarray}
where $E:=\frac{3\kv^2}{4M}-\frac{\gamma^2}{M}-\ii\epsilon$ is the
total non-relativistic energy; the amplitude $t^{(l)}_{d,tt}$ for
the $Nd_t\to Nd_t$-process; $t^{(l)}_{d,ts}$ for the $Nd_t\to
Nd_s$-process; $\kv$ the relative momentum of the incoming deuteron;
$\pv$ the off-shell momentum of the outgoing one, with $p=k$ the
on-shell point; and the deuteron propagator only at LO is:
\begin{equation}\label{eq:dprop}
D(q_0,\qv)=\frac{1}{\gamma-\sqrt{\frac{\qv^2}{4}-Mq_0-\ii\epsilon}}.
\end{equation}
The projected propagator of the exchanged particle on angular
momentum $l$ is given by:
\begin{equation}
  \label{eq:projectedNpropagator}
  \calK^{(l)}(E;q,p):=\frac{1}{2}\;\int\limits_{-1}^1\deint{}{x}
  \frac{P_l(x)}{p^2+q^2-ME+ pqx}
  =\frac{(-1)^l}{pq}\;Q_l\left(\frac{p^2+q^2-ME}{pq}\right)\;,
\end{equation}
where $Q_l(z)$, $l$th Legendre polynomial is:
\begin{equation}
  \label{eq:legendreQ}
   Q_l(z)=\half\int\limits_{-1}^1\deint{}{t}\frac{P_l(t)}{z-t}\;\;, \non
\end{equation}

(b) Lagrangian that includes the new term generated by the
two-derivative three-body force. Contrary to the terms without
derivatives, there are different, inequivalent three-body force
terms with two derivatives. But only one of them, $H_2$ is enhanced
over its naive dimensional estimate, mandating its inclusion at
N$^2$LO~\cite{4stooges,20}.

In that case, a unique solution exists in the $^2S_{1/2}$-channel
for each $\Lambda$ and vanishing three-body force, but no unique
limit as $\Lambda\to\infty$.  As long distance, phenomena must
however be insensitive to details of the short-distance physics (and
in particular of the regulator chosen), Bedaque et
al.~\cite{3stooges_boson,3stooges_doublet,4stooges,griesshammer}
showed that the system must be stabilized by introducing of a
three-body force in a form:
\begin{equation}
  \label{eq:calH}
   \calH(E;\Lambda)=
   \frac{2}{\Lambda^2}\sum\limits_{n=0}^\infty\;H_{2n}(\Lambda)\;
   \left(\frac{ME+\gamma_t^2}{\Lambda^2}\right)^n.
\end{equation}
which absorbs all dependence on the cutoff as
$\Lambda\to\infty$~\cite{4stooges}. At leading order,
$\gamma\ll\Lambda$, the essential observation are of order
$\mathcal{O}(Q/\Lambda)$ and are independent of energy (or momentum)
and hence can be made to vanish by only $H_0$ and not any of the
higher derivative three-body forces. At NLO, there is only a
perturbative change from the LO asymptotic of order
$\mathcal{O}(1/\Lambda^2)$(see ~\cite{4stooges}).

For N$^2$LO order, a new three-body force seems to be required. The
terms of order $(Q/\Lambda)^3$ is seems to be required, which
guarantees that the amplitude is cutoff independent up to
$(Q/\Lambda)^2$. The terms will be proportional to $ME$ arising from
expanding the kernel in powers of $q/\Lambda$ and
$\sqrt{ME}/\Lambda$ and a three-body force term which contains a
dependence on the external momenta $k^2$ and $ME$ can absorb them.
It means we need $H_2$ for calculations up to N$^2$LO. It has been
shown that the three body system at N$^2$LO can be renormalized
without the need for an energy dependent three-body force at this
order~\cite{platter-FBS40}. However, this work only discusses
three-nucleon scattering at cutoffs, up to 500 MeV, that are large
compared to the pion mass.

(c)  Lagrangian that describing currents, contributing to
$nd\rightarrow {^3H}\gamma$, which are not included in the previous
two Lagrangians.

The polarization can be characterized by the three-vector $\vec
\epsilon $, which satisfies the Lorentz condition: $\vec \epsilon
\cdot \vec k=0$, where $\hat{\vec k}$ is the unit vector along the
photon three-momentum. The sum over the photon polarizations is
given by:
$$\sum_{i=1,2}\epsilon_a^{(i)*}\epsilon_b^{(i)}=\delta_{ab}-\hat {k_a}\hat {k_b},$$
where the upper index $i$ numerates the two possible independent
polarization vectors of the photon.

For photo production processes or radiative capture, the photon can
be characterized by the smallest value of $j$. In this case, we can
easily write the corresponding combinations of polarization $\vec
\epsilon$ and unit vector $\hat{\vec k}$ for the photon states with
low multipolarity:
\begin{eqnarray*}
&\vec \epsilon&\to E1 ~\mbox{(electric~ dipole)},\\
&\vec \epsilon\times\hat {\vec k}&\to M1 ~\mbox{(magnetic~ dipole)},\\
&\epsilon_a\hat k_b+\epsilon_b\hat k_a \equiv E_{ab}&\to E2 ~\mbox{(electric~ quadrupole)},\\
&(\vec \epsilon\times\vec k)_a\hat k_b+(\vec \epsilon\times\vec
k)_b\hat k_a
\equiv M_{ab}&\to M2 ~\mbox{(magnetic~ quadrupole)}.\\
\end{eqnarray*}
These are the basic formulas for the construction of the matrix
elements for electromagnetic processes ( in near threshold
conditions). Main part of its contribution will be $E_1$ transitions
for photon energies below 20~MeV. If the ground state of $^3H$ is
thought to be an almost pure $^2S_{1/2}$ state, an electric-dipole
transition gives a $^2P_{1/2},_{3/2}$ final state.

Cross sections for transitions to final states with isospin $T =
3/2$ correspond to the three-body breakup while those to $T = 1/2$
are known to correspond predominantly to the two-body breakup. The
resulting matrix element for $E_1$ transition of $nd\rightarrow
{^3H}\gamma$ process, can be written as:
\begin{equation}
\left (\vec \epsilon \cdot\vec q\right ) ~\left (t^\dagger
\sigma_2\vec\sigma~\cdot\vec D^*N\right ),
\end{equation}
where $N$, $t$, $\vec q$ and $\vec D$ are the 2-component spinors of
initial nucleon, final $^3H$, the unit vector along the 3-momentum
of the colliding nucleons and the polarization vector of the
deuteron, respectively. The spin structure of the matrix elements
for $N+d$ radiative capture is complicate also for the low energy
interaction, as we have here 3 independent multipole transitions:
$j^{P}=\displaystyle\frac{1}{2}^+\to M1$ and $
j^{P}=\displaystyle\frac{3}{2}^+\to M_1$ and $E_2$. The following
parametrization of the corresponding contributions to the matrix
element are:
$$i(t^\dagger N)(\vec D\cdot\vec{\epsilon^*}\times\vec k),$$
\begin{equation}
(t^\dagger\sigma_a N)(\vec D\times [\vec{ \epsilon^*}\times\vec
k])_a, \label{eq:as7}
\end{equation}
$$t^\dagger(\vec\sigma\cdot\vec{\epsilon^*}~\vec D\cdot\vec k+\vec\sigma\cdot
\vec k~ \vec D\cdot\vec \epsilon^*)N,$$
 The first two structures in
(\ref{eq:as7}) are correspond to the $M_1$ radiation.

The $E_1$ and $M_1$ amplitudes receive contributions from the
electric and magnetic moments of the nucleon and from four-nucleon
operators coupling to the magnetic field, which are described by the
two-body currents contributing to $nd\rightarrow {^3H}\gamma$ or
${^3H}\gamma\rightarrow nd$~\cite{Rupak}:

\begin{eqnarray}\label{eq:M1}
\mathcal{L}_{EM}&=&\frac{e}{2M_N}N^\dagger(k_0+k_1
\tau^3){\bf{\sigma.B}}N +e\,L_{B}(N^T P_j N)^\dagger(N^T
\overline{P}_3 N){\bf B}_j\nonumber\\
&{}& +\frac{1}{2}e\,L_{E}(N^T(\stackrel{\leftarrow}{\bf D}_j
P_a^{(1)}-P_a^{(1)} \stackrel{\rightarrow}{\bf D}_j) N)^\dagger(N^T
P_a N) {\bf E}_j+H.C. \end{eqnarray}

where ${\bf E}$, ${\bf B}$ and $P_a$ are the electric, magnetic
fields and the projector for the $^1S_0$ channel, respectively. The
covariant derivative is:
\begin{equation}
D^\mu=\partial^\mu +ie\frac{1+\tau_3}{2} A^\mu\nonumber\ ,
\end{equation}
and $P_i^{(1)}$ are the spin-isospin projectors for the isotriplet,
spintriplet channel
\begin{equation}
P_i^{(1)}\equiv\frac{1}{\sqrt{8}}\sigma_2\sigma_i\ \tau_2\tau_3\ ,
\hspace{.2in}{\rm Tr}[P_i^{(1)}P_j^{(1)}]=\frac{1}{2}\delta_{ij}.
\end{equation}

The $L_{B}$ and $L_{E}$ are determined from the two-body
neutron-proton radiative capture and photodisintegration of the
deuteron data, respectively, see~\cite{Rupak} for more details.

The diagrams of $E_1$ contribution is shown in fig.~1.  This figure
shows simple inclusion of photon to the Faddeev equation
contribution to $E_1$ amplitude. One of the other possible diagrams
which could be contributed in the calculation is the insertion
photon to exchanged nucleon(middle diagram in fig.~1).  This diagram
gives a tiny contribution when the photon couples electrically to
exchanged nucleon, i.e. via $\vec{P} \cdot \vec{A}$. This is due to
the fact that the initial and final state in this process are nearly
orthogonal.

A higher dimension operator coupling an $E_1$ photon to a
$^3S_1$-dibaryon and a p-wave two nucleon final state is suppressed
by $Q^3$ in the power counting. This calculation involve only  $E_1$
contribution. Meson exchange currents is another diagram to be
incorporated and it is seen that electric quadrupole $E_2$
contributions are needed to achieve the better agreement between
theory and experiment, specially at high energies.  For higher
energies, the present results are not complete and calculations with
$E_2$ contribution must be carried out.

The $M_1$ amplitude receives contributions from the magnetic moments
of the nucleon and dibaryon operators coupling to the magnetic
field. Fig.~2 shows the diagrams for adding photon to the Faddeev
equation up to N$^2$LO. There is no interaction of $H_0$ with a
photon, because $H_0$ has no derivatives, so it is not affected by
the minimal substitution $\vec{P}\rightarrow \vec{P} - e \vec{A}$.
Contribution from the photon couples to the
three-nucleon field ($H_2$) is not negligible(fig.~2) and should be
calculated up to order of calculation. This coupling comes from the
kinetic energy insertion of the three-nucleon field.

\begin{figure}[!htp]
\begin{center}
 \includegraphics*[width=.6\textwidth]{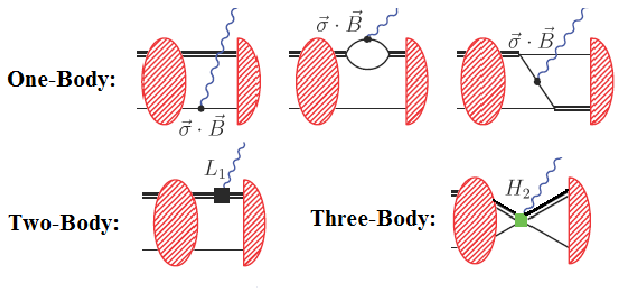}
 \vspace*{-0pt}
\caption{ The Faddeev equation for $Nd$-scattering and some diagrams
for adding photon-interaction to the Faddeev equation contribution
to $M_1$ amplitude. The small circles show magnetic photon
interaction. Vertices for photon interaction to two- and three-body
system have been shown by $L_1$ and $H_2$, respectivly. Remaining
notation as in fig.~\ref{fig1}.} \label{fig2}
\end{center}
\end{figure}

\section{ Results and discussion}
\label{section:comparison}

 The numerical calculations are done using $B=2.225$ MeV
($\gamma_d=45.7066$ MeV)for the $NN$ triplet channel (deuteron
binding momentum), $a_s=-23.714$fm for the $NN$ singlet channel an
$^1S_0$ scattering length.  At LO and NLO, $H_0$ is the only
three-body force entering and $H_2$is also required at N$^2$LO. The
$H_0$ and $H_2$ three-body parameters are fixed from the
${}^2\mathrm{S}_\frac{1}{2}$ scattering length $a_3=(0.65\pm0.04)$
fm and the triton binding energy $B_3=8.48$ MeV, respectively. The
$L_{B}$ and $L_{E}$ are obtained by fixing at leading non-vanishing
order by the thermal neutron-proton radiative capture and deuteron
photodisintegration cross sections, respectively~\cite{Rupak}.

In order to compute a solution, one introduces a cutoff $\Lambda$
which is nonphysical and thus not to be confused with the breakdown
scale $\LambdaNoPion$ of the EFT. We show cutoff variation of cross
section between $\Lambda=150$ MeV and $\Lambda=500$ MeV as a
function of the center of mass momentum is shown in fig.~3. A
numerical investigation of the Faddeev equation confirms these
findings. One can see smooth slope and does however change
significantly from order to order, because the dominant correction
is $~(\gamma/\Lambda)^n$ and for these energy range at every order.
These variations are nearly independent of variation of momentum. We
confirm that, the cutoff variation decreases steadily as we increase
the order of the calculation and it is of the order of
$(k/\Lambda)^n, (\gamma/\Lambda)^n$, where $n$ is the order of the
calculation and $\Lambda=150$ MeV is the smallest cutoff used. The
other errors due to increasing momentum also appear in our
calculation in low energy but these errors decrease when the order
of calculation is increased up to N$^2$LO.
\begin{figure}[!htp]
\begin{center}
\includegraphics[width=.6\linewidth,clip=true]{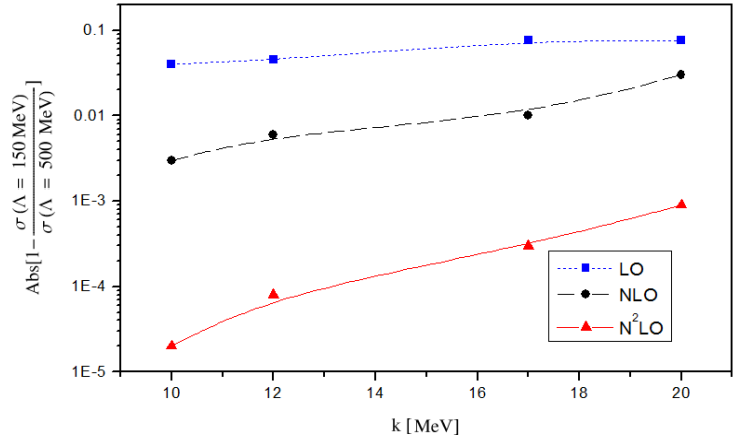}
\vspace*{-0pt} \label{fig3} \caption{Curve of  the cutoff variation
of  cross section up to N$^2$LO is shown between  $\Lambda=150$ MeV
and $\Lambda=500$ MeV. The short dashed, long dashed  and solid line
correspond to LO, NLO and N$^2$LO, respectively.}
\end{center}
\end{figure}

\begin{table}[!htp]
\begin{center}
\begin{tabular}{c||c|c|c|c|c|c|c|c c}
\hline $E_\gamma$  & $\sigma(M_1)$ & $\sigma(E_1)$ &
$\sigma(M_1+E_1)$ &
$\sigma(M_1)$ & $\sigma(M_1+E_1)$& $\sigma(M_1)$ & $\sigma(M_1+E_1)$ & Exp. \\
$(MeV)$ & $(LO)$ & $(LO)$ & $(LO)$ &
$(NLO)$ & $(NLO)$& $(N^2LO)$ &  $(N^2LO)$ & $(mb)$\\
\hline \hline
$11.54$ & $0.039$ & $0.782$ & $0.821 $ & $0.054$ & $0.833 $ & $0.063$ & $0.845 $ & $0.84\pm0.05$\\
$11.92$ & $0.037$ & $0.800$ & $0.837 $ & $0.050$ & $0.855 $ & $0.060$ & $0.860 $ & $0.95\pm0.05$\\
$12.55$ & $0.034$ & $0.820$ & $0.854 $ & $0.045$ & $0.870 $ & $0.055$ & $0.875 $ & $0.96\pm0.06$\\
$13.04$ & $0.030$ & $0.830$ & $0.860 $ & $0.041$ & $0.876 $ & $0.049$ & $0.880 $ & $0.86\pm0.05$\\
$13.92$ & $0.027$ & $0.831$ & $0.858 $ & $0.038$ & $0.874 $ & $0.047$ & $0.878 $ & $0.79\pm0.06$\\
$14.92$ & $0.026$ & $0.827$ & $0.853 $ & $0.036$ & $0.868 $ & $0.045$ & $0.872 $ & $0.71\pm0.06$\\
$15.92$ & $0.023$ & $0.820$ & $0.843 $ & $0.033$ & $0.858 $ & $0.042$ & $0.862 $ & $0.61\pm0.05$\\
$16.80$ & $0.020$ & $0.810$ & $0.830 $ & $0.030$ & $0.845 $ & $0.038$ & $0.848 $ & $0.85\pm0.06$\\
$17.66$ & $0.018$ & $0.796$ & $0.814 $ & $0.027$ & $0.828 $ & $0.035$ & $0.835 $ & $0.74\pm0.07$\\
$18.67$ & $0.015$ & $0.786$ & $0.801 $ & $0.024$ & $0.815 $ & $0.032$ & $0.818 $ & $0.66\pm0.07$\\
$19.16$ & $0.014$ & $0.778$ & $0.792 $ & $0.022$ & $0.805 $ & $0.030$ & $0.808 $ & $0.66\pm0.07$\\
$19.72$ & $0.013$ & $0.770$ & $0.783 $ & $0.021$ & $0.796 $ & $0.029$ & $0.799 $ & $0.71\pm0.11$\\
\hline
\end{tabular}
\caption{ The cross section for $\gamma ^3H\rightarrow nd$ in
millibarns computed in pionless EFT up to N$^2$LO. The pionless EFT
cross section is comprised of the $E_1$ amplitude computed to LO and
the M1 amplitude computed to LO, NLO and N$^2$LO. The last column
shows the experimental date from~\cite{Faul}.}
\end{center}
\end{table}

The breakdown of pionless EFT in calculation is estimated to be
corresponds to photon energies of $m_{\pi}^2/M \approx$ 20 MeV. The
photodisintegration reaction $\gamma {^3H}\rightarrow nd$ results in
the low photon energy range(0-20~MeV) are shown in table 1. It is
well known that this reaction will be occurred predominantly via
electric and magnetic, specially electric dipole transition. Our
obtained results for electric dipole transition ($E_1$) show higher
contribution in comparison with magnetic dipole transition ($M_1$)
at these energies. The ($E_1$) contribution is calculated for  LO
and the next leading order is beyond  N$^2$LO. The ($M_1$)
contributions are calculated for  LO, NLO and N$^2$LO. The
calculated cross section shows sharply rising from threshold to
maximum about 0.88 mb at $\sim$13 MeV and decreasing only slightly
to about 0.81 mb at $\sim$19 MeV, in corresponding to experimental
results~\cite{Faul}.

\begin{table}[!htp]
\begin{center}
\begin{tabular}{c||c|c}
\hline  Type of Formulation & $E_{\gamma}=12$ [MeV] & $E_{\gamma}=40$ [MeV]\\
\hline\hline
AV18-Siegert             & $1.056$ & $0.168$ \\
AV18-MEC                 & $0.949$ & $0.155$  \\
CD Bonn2000-Siegert      & $0.980$ & $0.169$  \\
\hline
AV18+UrbanaIX-Siegert    & $0.882$ & $0.180$  \\
AV18+UrbanaIX-MEC        & $0.915$ & $0.169$  \\
CDBonn2000+TM'-Siegert   & $0.889$ & $0.176$  \\
EFT(N$^2$LO)   & $0.874$ & $0.240$  \\
\hline
\end{tabular}
\caption{The total cross section results(in mb) for two-body
photodisintegration of $^3H$ of different models-dependent based on
Faddeev approach~\cite{Skibinski}. The last row shows
model-independent EFT result up to N$^2$LO.}
\end{center}
\end{table}

In order to provide information on the dependence of the cross
section calculation on the choice of models, forces and currents the
results for the total two-body photodisintegration cross section of
$^3$H  have been displayed at two energies in table 2. At 12 MeV,
one can see a good agreement between the results of AV18 and UrbIX
with three-nucleon forces and pionless EFT result up to N$^2$LO. It
is also found, the two-body photodisintegration of the triton cross
section is dominated by $E_1$ transition.

Comparison of the different models, show that the results of
pionless EFT up to N$^2$LO are in good agreement with the other
modern nucleon-nucleon potentials with three-nucleon forces. Table 2
illustrates also that the result with explicit MEC differs rather
substantially from that with the one-body current and implicit MEC
via Siegert's theorem as well as EFT results. However, precise data
would be required in order to reach a better conclusion.

\begin{figure}[!htp]
\begin{center}
\includegraphics[width=0.5\linewidth,clip=true]{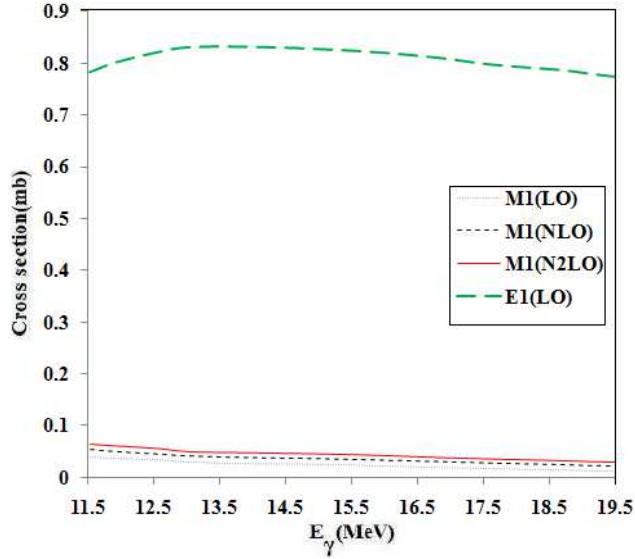}
\vspace*{-0pt} \caption{$^3$H photo-dissociation cross section
$\sigma$ (mb) for center of mass energy E (MeV). Dot, dash and solid
curve are LO, NLO, and N$^2$LO $M_1$ transitions,and long-dash is LO
$E_1$ transition.} \label{fig4}
\end{center}
\end{figure}

The leading contributions from each transitions are shown in fig.~4.
One can see that the $M_1$ transition is small in comparisons to
$E_1 $ transitions and as a results the $E_2$ transitions can be
ignored for calculation in the energy range of our interest. A
comparison between the low energy cross section computed with
pionless EFT and the experimental data as well as the potential
model calculation, can be seen in fig.~5. A detailed and very
illuminating discussion of the experiments that contributed to these
data points can be found in ref.~\cite{Faul}. One finds a rather
good agreement between pionless EFT theory and experiment at low
energy. Since the experimental situation is not settled, a good
definite comparison between theory and experiment cannot be made. It
is due to the fact that the two-body break up data are rather
scattered and there is also limited experimental information on the
three-body break up.

We also compared the obtained total $^3$H cross section with the
other modern nucleon-nucleon AV18/UrbanaIX ~\cite{Skibinski}, in
fig.~5. One finds a rather good agreement between EFT results up to
N$^2$LO and the experimental data, as well as, potential models that
the three-nucleon forces improve their consistency with the
experiments. On the other hand, due to the insufficient precision of
the experimental data a more conclusive comparison between theory
and experiment cannot be made.  As another result, we leave the
consistency of our results for energies higher than $\sim$ 20 MeV.
This is in agreement with the breakdown energy scale of pionless EFT
($\sim$ 20 MeV). Effects of others multipoles, retardation, tensor
correlations, and subnuclear currents should have more and more
influence at higher energies.

\begin{figure}[!htp]
\begin{center}
\includegraphics[width=0.7\linewidth,clip=true]{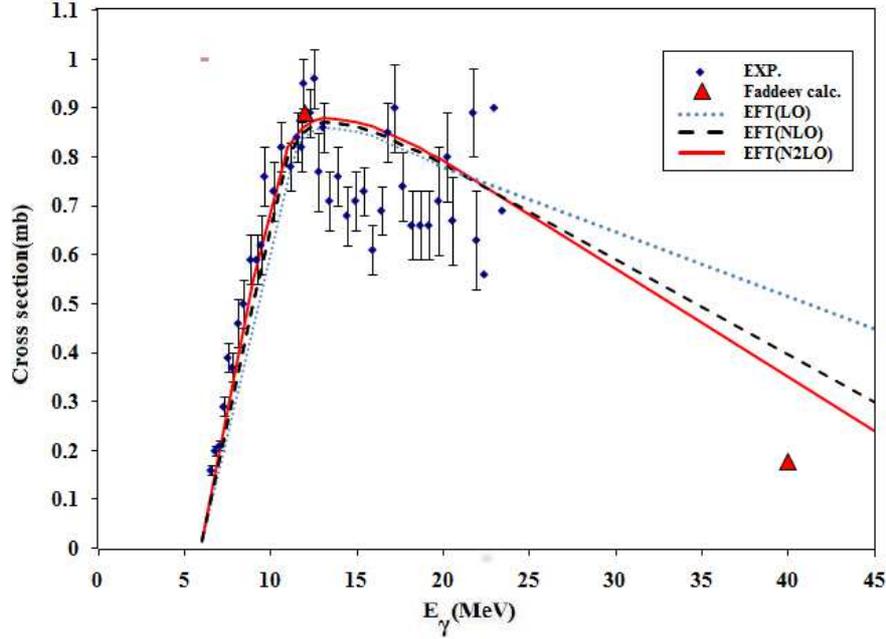}
\vspace*{-0pt} \caption{The cross section for $\gamma
{^3H}\rightarrow nd$ near threshold, as a function of the incident
photon energy in MeV. The solid curve corresponds to the cross
section computed in pionless EFT up to N$^2$LO.  The dotes with
error bar correspond to the experimental date from~\cite{Faul} and
triangle points show the Faddeev calculation results for cross
sections of the $^3$H photodisintegration~\cite{Skibinski}. The
short dashed, long dashed and solid line correspond to LO, NLO and
N$^2$LO effective field theory calculation, respectively.}
\label{fig5}
\end{center}
\end{figure}

Our result using EFT is model independent and universal, any model
with the same input must, within the accuracy of our calculation,
lead to the same result. Due to the large error bars in experimental
data, it is not possible to draw further conclusions.

\section{Summery and conclusion}
\label{section:conclusion}

We calculated the cross section of photo-dissociation of
$\gamma{^3H}\rightarrow nd$. We applied EFT to find numerical
results for this process. The results are compared with, AV18
together with the Urbana~IX three-nucleon forces or CD Bonn with TM
three-body forces, high precision other nuclear force models. Our
results are in good agreement with the other calculations and
experimental data. Theoretical predictions show the maxima sharply
rising from threshold to maximum about 0.88 mb at $\sim$ 13 MeV and
decreasing only slightly to about 0.81 mb at $\sim$19 MeV, in
corresponding to experimental results.

The results are quantitatively supported by available data and
converge order by order in low energy expansion. The results are
also cutoff independent when the order of calculation is increased
up to N$^2$LO.  High quality experimental data would be very helpful
more strongly to challenge theory at low energies.

\section{Acknowledgments}
The authors would like to thank Harald W. Grie\ss hammer for useful
and valuable comments.

\end{document}